\documentclass[]{spie}  

 \usepackage{chemfig}
\usepackage[version=4]{mhchem}
\usepackage{indentfirst}
\usepackage{geometry}
\usepackage{mathabx}
\usepackage[export]{adjustbox}
\usepackage{relsize}
\usepackage{graphics}
\usepackage{graphicx}
\usepackage{caption}
\usepackage[font=small,labelfont=bf]{caption}
\usepackage{float}
\usepackage{rotating}
\usepackage{lscape}
\usepackage{tikz}
\usepackage{sidecap}
\usepackage{amsmath}
\usepackage{bm}
\usepackage{comment}
\usepackage[colorlinks=true, allcolors=blue]{hyperref}

\title{Simulating the Study of Exoplanets Using Photonic Spectrographs}

\author[a]{Marcos Perez$^*$}
\author[a]{Pradip Gatkine$^{*}$}
\author[a]{Nemanja Jovanovic}
\author[b]{Jeffrey Jewell}
\author[b]{J. Kent Wallace}
\author[a,b]{Dimitri Mawet}
\affil[a]{California Institute of Technology, 1200 E. California Blvd, Pasadena, United States}
\affil[b]{Jet Propulsion Laboratory, 4800 Oak Grove Drive, La Cañada Flintridge, United States}

\authorinfo{Further author information: (Send correspondence to M. Perez or P. Gatkine)\\ E-mail: marcos.aaron.perez@gmail.com, pgatkine@caltech.edu}
\begin{document} 
\maketitle

\def\thefootnote{*}\footnotetext{These authors contributed equally to this work}\def\thefootnote{\arabic{footnote}}

\begin{abstract}
Photonic spectrographs offer a highly miniaturized, flexible, and stable on-chip solution for astronomical spectroscopy and can be used for various science cases such as determining the atmospheric composition of exoplanets to understand their habitability, formation, and evolution. Arrayed Waveguide Gratings (AWGs) have shown the best promise to be used as an astrophotonic spectrograph. We developed a publically-available tool to conduct a preliminary examination of the capability of the AWGs in spectrally resolving exoplanet atmospheres. We derived the Line-Spread-Function (LSF) as a function of wavelength and the Full-Width-at-Half-Maximum (FWHM) of the LSF as a function of spectral line width to evaluate the response of a discretely- and continuously-sampled low-resolution AWG (R $\sim$ 1000). We observed that the LSF has minimal wavelength dependence ($\sim$5\%), irrespective of the  offset with respect to the center-wavelengths of the AWG channels, contrary to the previous assumptions. We further confirmed that the observed FWHM scales linearly with the emission line width. Finally, we present simulated extraction of a sample molecular absorption spectrum with the  discretely- and continuously-sampled low-resolution AWGs. From this, we show that while the discrete AWG matches its expected resolving power, the continuous AWG spectrograph can, in principle, achieve an effective resolution significantly greater ($\sim$ 2x) than the discrete AWG. This detailed  examination of the AWGs will be foundational for future deployment of AWG spectrographs for astronomical science cases such as exoplanet atmospheres. 
\end{abstract}

\keywords{Astrophotonic spectrographs,  Arrayed Waveguide Gratings,  On-chip, Line Spread Function, Exoplanets}

\section{INTRODUCTION}
\label{sec:intro}  
\subsection{Background and Motivation}
\subsubsection{Exoplanets}
As of September 2021, over 4,521 planets have been discovered orbiting a star other than the Sun, known as exoplanets \cite{akeson_nasa_2013}.  In order to better understand our place in the universe, and to search for extraterrestrial life beyond the Earth, it is of key importance that it is determined which exoplanets could host life. Determining the atmospheric composition of exoplanets is necessary for understanding their habitability, planet formation, and evolution.\par
There are two primary methods of conducting spectroscopy of exoplanet atmospheres: spectroscopy of exoplanets in transit (passing through the line of sight between Earth and their host star) and spectroscopy of directly imaged exoplanets.
\cite{kreidberg_exoplanet_2017}. By using the world's largest, most advanced telescopes it may be possible to measure the presence of compounds that may indicate a potential environment for life ($\chemfig{CH_4}, \chemfig{O_2}, \chemfig{H_2O}$) \cite{seager_exoplanet_2013}.  
Historically, this has been a challenging endeavour, but by using high resolution spectrographs, and high contrast exoplanet imaging technologies, it is possible to separate spectra associated with the star from spectra associated with the planet's atmosphere \cite{kreidberg_exoplanet_2017}\cite{gatkine_astrophotonic_2019}. The Keck Planet Imager and Characterizer (KPIC) on the W.M. Keck II Telescope is an instrument suite that has demonstrated high resolution spectroscopy of the atmospheres of gas giant exoplanets in the infrared \cite{wang2021high}\cite{mawet2021enabling}.

\subsubsection{Photonic Spectrograph: Arrayed Waveguide Grating (AWG)}

Currently, the Keck Planet Imager and Characterizer (KPIC) employs a conventional bulk-optics spectrograph, NIRSPEC, which is 1.0 m $\times$ 1.5 m $\times$ 0.7 m large \cite{mclean1998design}.
Photonic spectrographs offer an on-chip and extremely compact solution that is suitable for a KPIC-like setup\cite{gatkine2019astrophotonic}. In addition, the guiding the light in single-mode waveguides on photonic chips grants unique capabilities such as spectral filtering (eg: using waveguide Bragg gratings\cite{zhu2016arbitrary} to filter the narrow OH-emission lines in the atmosphere), spatial filtering (eg: using on-chip nulling interferometers to suppress starlight and thus enhance the star-planet contrast for exoplanet spectroscopy\cite{norris2020first, gatkine_astrophotonic_2019}), and polarization filtering (by leveraging the polarization dependence of the waveguides). A photonic spectrograph can integrate one or more of these capabilities with spectroscopy thanks to the single-mode propagation and hence, offer access to unique science cases.  

\begin{figure}[h]
\centering
\includegraphics[width=1.0\textwidth]{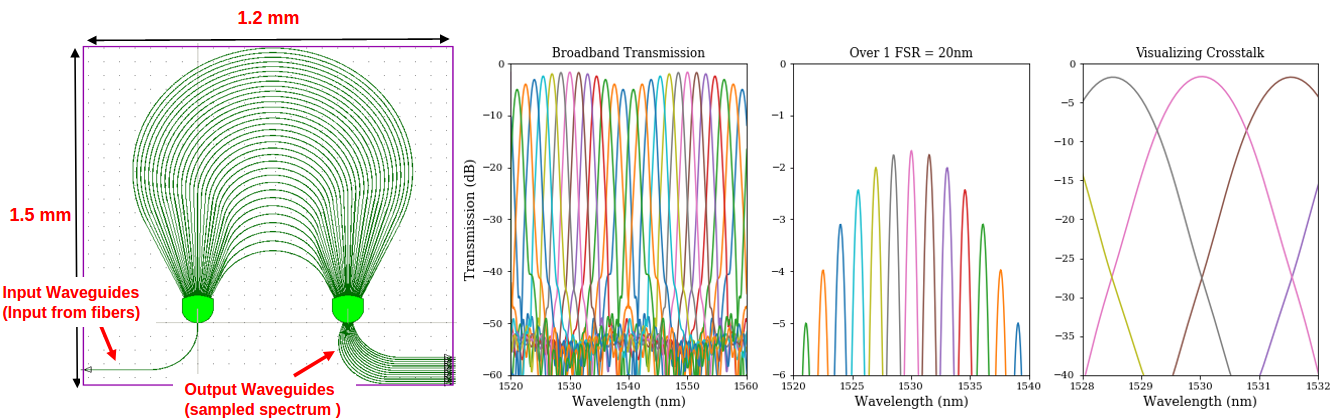}
\captionsetup{width=.85\linewidth}
\captionof{figure}{Left: CAD of the AWG, Right: Transmission response of the AWG in the form of Transmission in decibels (dB) as a function of wavelength.}\hfill
\label{fig:figure1}
\end{figure}

In this paper, we conduct preliminary simulations to examine the capability of photonic spectrographs in characterizing and spectrally resolving the atmospheres of exoplanets \cite{morris_keck_2020}.  So far, the Arrayed Waveguide Gratings (AWG) has been the best photonic architecture for building a compact diffraction-limited astrophotonic spectrograph \cite{cvetojevic2012first,  gatkine2016development, gatkine2019astrophotonic, stoll2020performance}.
The AWG-based photonic spectrograph intakes light from single-mode optical fibers. Light is then directed into an array of waveguides with different path lengths  leading to a constructive interference of different wavelengths at different locations along its focal plane (the Rowland curvature), allowing the AWG to create a spectrum which is then sent to a detector to be recorded, as shown in Fig. \ref{fig:figure1}. Each output waveguide in Fig. \ref{fig:figure1} corresponds to a discrete spectral channel. 

We seek to further the development of photonic spectrographs by running simulations to systematically gauge their capabilities to reliably acquire high-quality exoplanet spectra using large telescopes. The advent of photonic spectrographs uniquely offers a drastic reduction in the size and mass of spectrographs as well as flexibility in coherently manipulating the light \cite{gatkine_astrophotonic_2019}. These advantages of photonic spectrographs make them well-suited for existing ground-based observatories, and future space-based telescopes \cite{Gatkine2019State}.

In the future, we will  simulate the performance of AWG spectrographs on KPIC for measuring the chemical composition of exoplanet atmospheres.\par

\section{LINE SPREAD FUNCTION (LSF)}
The first step is to characterize how the AWG transmits different wavelengths of light. Even if the AWG spectrograph is illuminated by a narrow emission line (much narrower than the spectral channel), multiple spectral channels will still receive the photons since each spectral channel has a finite non-zero response at each wavelength (due to the inherent properties of the spectrograph such as side lobes and phase errors\cite{gatkine2021potential}). This ‘scattering' is called the line spread function (LSF). We aim to characterize the wavelength dependence of the LSF of an AWG by simulating the transmission of a narrow emission line positioned at various wavelengths in within the AWG's free spectral range. We developed a Python tool to conduct these simulations and \textbf{made the repository publicly accessible\footnote{https://github.com/MarcosP7635/2021\_SURF\_Marcos\_Perez} on Github} \cite{perez_marcosp76352021_surf_marcos_perez_2021}.

\subsection{Design}

\paragraph{AWG Design:} As a first step, we simulate a low-resolution AWG of R $\sim$ 1000 to examine the overall behavior of an AWG spectrograph. Note that, we define R = $\lambda/\delta\lambda$ with $\delta\lambda$ = channel spacing of the discrete AWG. Such low-resolution AWG is suitable for spectroscopy of faint targets such as exoplanets. We have considered two variants of the AWG, first with discrete output waveguides (i.e. sampling the spectrum into discrete output channels), and a second variant (of the same AWG) with a continuous sampling of the spectrum. The continuous-sampling AWG is achieved by simply removing the output waveguides from the discrete AWG and directly imaging the free-propagation region (FPR) on a detector \cite{cvetojevic2012developing, gatkine2017arrayed}. A CAD of the low-res AWG and its transmission response are shown in Fig \ref{fig:figure1}. The channel spacing and the free spectral range (FSR) of the AWG are 1.5 nm and 20 nm, respectively. The crosstalk for the discrete output AWG is -25 dB. The average full-width at half-maximum (FWHM) of each channel's wavelength response (also called the 3-dB width) is 1 nm. The scheme for examining the performance of this AWG as a spectrograph is detailed below and will be expanded to high-resolution AWGs (R $>$ 10,000) in the future. 

\subsection{Discrete AWG simulation}\label{subsec:discrete_AWG}
First, we consider the discrete-channel AWG to examine the spectrograph properties. The transmission response of the AWG, as shown in Fig. \ref{fig:figure1}, was obtained by simulating the AWG with beam propagation method using the Rsoft software \cite{BeamPROP}. 

\begin{figure}[h!]
\centering
    \includegraphics[width=\textwidth]{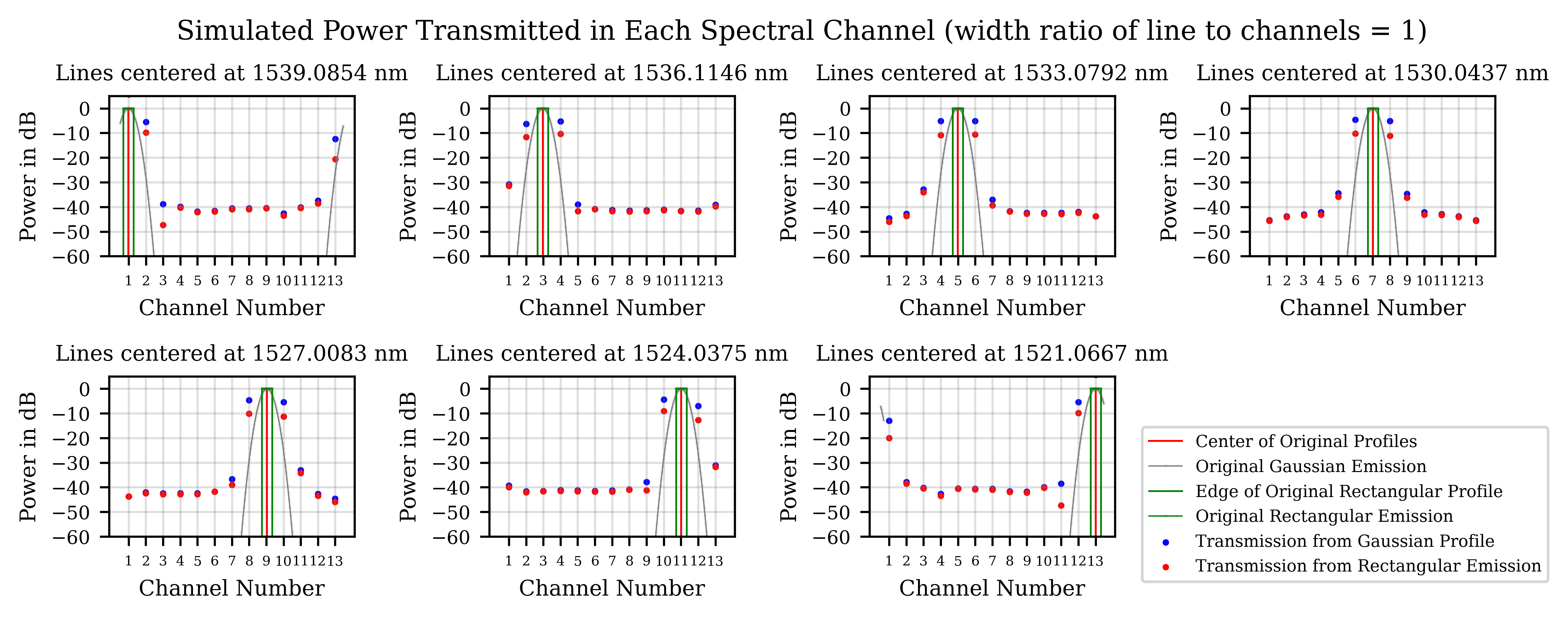}
    \caption{The simulated power measured by the detector after being dispersed by the Arrayed Waveguide Grating (AWG) when observing an emission line at the transmission peak of a spectral channel. We show both the \textbf{transmission for Gaussian (blue points) and rectangular (red points) input emission lines}. Each rectangular profile has a height of 1 (in power units) and a width equal to the average 3dB width of each AWG channel (1 nm). Each Gaussian profile has a peak of 1 and FWHM equal the average 3dB width of each channel (1 nm). Each emission line is centered at the peak of an odd numbered channel (channels 1, 3, 5, 7, 9, 11, and 13, respectively). Note that power in decibels (dB)  = 
    $\mathlarger{10\cdot\log_{10}\frac{P_{transmitted}}{P_{source}}}$ where $P_{transmitted}$ is the power transmitted by the spectrograph in a particular channel and $P_{source}$ is the original total power of the incident light (here taken to be 1 in arbitrary power units).}
    \label{fig:LSF_peaks_discrete}
\end{figure}

In order to measure the line-spread function, we simulated the transmission of a narrow emission line centered at the transmission peak of each spectral channel and calculated the power measured in each output waveguide.
The output power for each channel for various emission lines are shown in Fig. \ref{fig:LSF_peaks_discrete}. For each plot in Fig. \ref{fig:LSF_peaks_discrete}, we show the transmission from an individual emission line. We simulated both a rectangular emission line (in red points) and a Gaussian emission line (in blue points). Each rectangular profile has a height of 1 unit and a width equal to the average 3-dB width of each channel (= 1 nm).  Each Gaussian profile has a peak height of 1 unit and an FWHM equal to the average 3-dB width of each channel (= 1 nm).  

Notably, the shape of the LSF appears to change as we shift the emission line though the peak-wavelengths of different channels of the AWG (Fig. \ref{fig:LSF_peaks_discrete}). The LSF is a symmetric Gaussian-like curve for wavelengths around the center of a given spectral order (corresponding to channels 4 to 10) and become asymmetric towards the edges of the spectral order (corresponding to channels 1-3 and 11-13). This effect arises is due to the asymmetry in the illumination pattern of the arrayed waveguides as seen from the output waveguides at the edges (as opposed to the central output waveguides). The LSF from the rectangular emission line is clearly narrower compared to the LSF for Gaussian emission line, which is primarily due to the larger width of the Gaussian emission line. This effect is discussed in more detail in Section \ref{subsec:FWHM}.

\begin{figure}[h!]
\centering
    \includegraphics[width=0.9\textwidth]{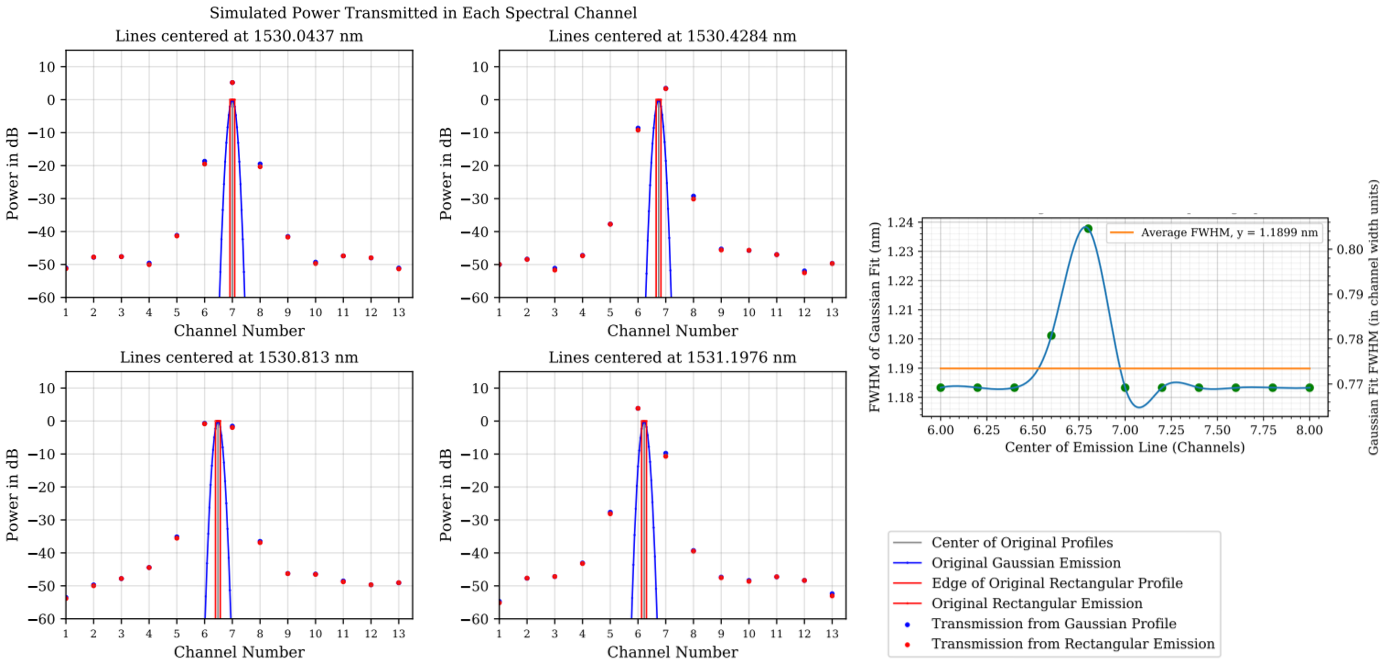}\\
    \caption{\textbf{Left and center:} The simulated power measured in the discrete output channels of the discrete AWG when emission lines of different profiles are input to the AWG. Each rectangular profile has a height of 1 and a width of  0.31 nm (= 1/5 channel spacing). Each Gaussian profile has a peak of 1 and FWHM = 0.31 nm. When only one of the emission line's transmission is visible on the plot, the rectangular and Gaussian emission line have approximately the same transmission.
    \textbf{Right: } The FWHM of the LSFs derived at various channel-offset values is shown, in the range of channel 6 to channel 8. The FWHM of the LSF at each offset was obtained by fitting a Gaussian to the resulting AWG power distribution, when the emission line is placed at the corresponding offset wavelength. It can be clearly seen that while the average FWHM for the LSF is around 1.19 nm, the FWHM varies by only $\sim$5\%, depending on the channel offset (and hence, the wavelength).}
    \label{fig:LSF_offset_discrete}
\end{figure}

\begin{figure}[h!]
\centering
        \includegraphics[width=0.85\textwidth]{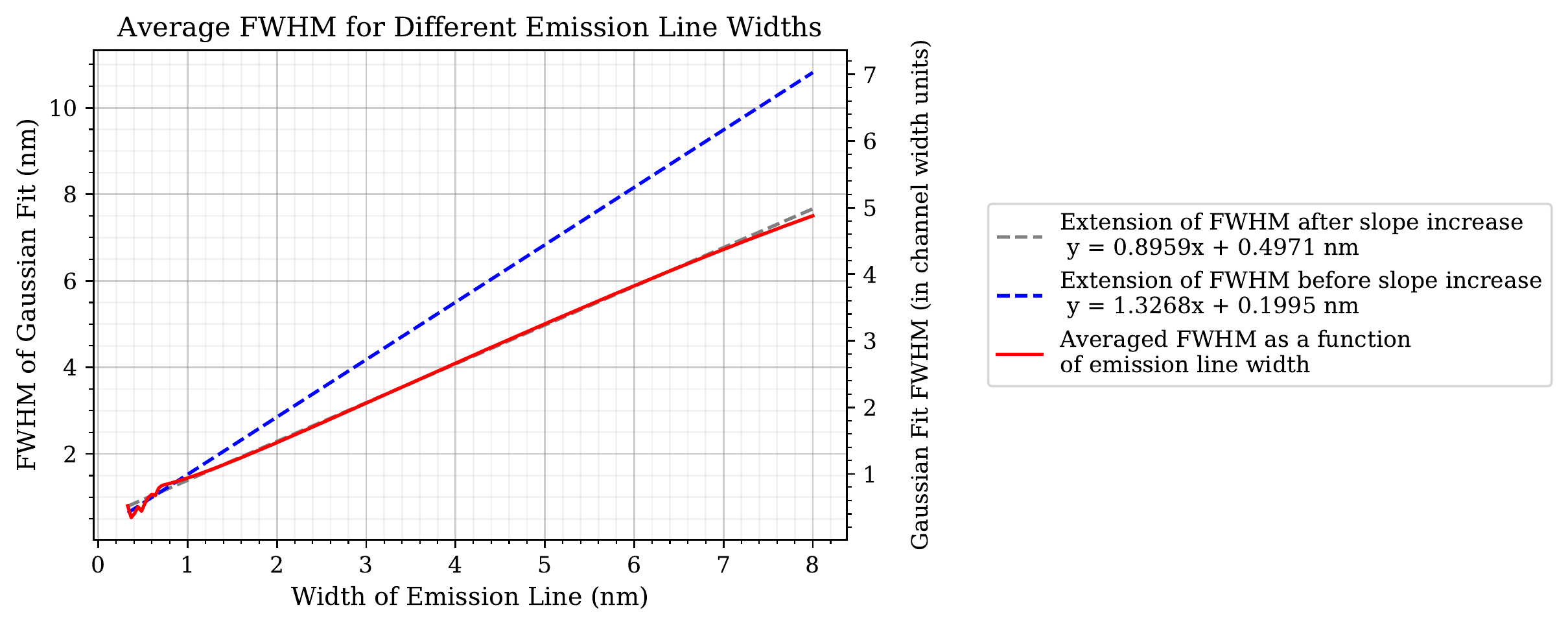}
        \caption{By minimizing the least square of the residuals for Gaussian fits of the LSFs for 2,200 different emission lines, we have calculated the FWHM of the photonic spectrograph as a function of emission line width (FWHM). Each red point is the average FWHM of the LSF for 11 emission lines of the same width but centered at different points between the peaks of channels 6 and 8.}
        \label{fig:FWHM_vs_em_width_discrete}
\end{figure}

The discrete AWG transmission response for a given channel varies sharply as a function of wavelength. For instance, there is 10 dB variation in the transmission as the wavelength changes from 1550 to 1549 nm for the magenta channel, as seen in panel 4 of Fig. \ref{fig:figure1}. Given such sharp changes, it is important to examine the LSF not only at the peak wavelengths of the channels, but also at various offset wavelengths (i.e. in-between the channel peaks). Following a similar procedure are above, we calculated the output power in AWG channels for emission lines placed at various wavelength offsets. We used 0.31 nm (= 1/5 $\times$ channel spacing) as the  width of rectangular emission line and FWHM of Gaussian line, instead of 1 nm to ensure that the line remains unresolved.    In Fig. \ref{fig:LSF_offset_discrete}, we show the AWG power distribution (and hence, the LSF) corresponding to emission line at the peak wavelength of channel 7 (1530 nm) as well as at the offsets of  25\% (1530.4 nm), 50\% (1530.8 nm), and 75\% (1531.2 nm) between channels 7 and 6.  By fitting a Gaussian to the resulting AWG power distribution, we obtained the FWHM of the LSFs. The FWHM for various offset values are shown in the rightmost panel of Fig. \ref{fig:LSF_offset_discrete}. It can be clearly seen that while the average FWHM is around 1.19 nm, the FWHM varies by $\sim$5\% within the range of 2 channels, depending on the offset from the channel peak wavelength. Thus, we conclude that the AWG LSF is mildly ($\sim 5\%$) wavelength-dependent. In Fig. \ref{fig:FWHM_vs_em_width_discrete}, we plot the FWHM of the observed (Gaussian-like) profile with the AWG against the FWHM of 2200 different emission lines input to the AWG (one line at a time). It is clear from the figure that the observed width scales linearly with the width of the source emission line, which confirms that there are no non-linear artifacts when we use the  AWG as a spectrograph, as expected. The small non-linearity is only seen when the emission line is unresolved (FWHM $<$ 1 nm), which is expected in a spectrograph.

\subsection{Continuous AWG}\label{subsec:continuous_AWG}
The continuous AWG is realized by removing the discrete output waveguides from the AWG, and thus allowing a continuous sampling of the spectrum by directly imaging the free propagation region (FPR, i.e. the focal plane) of the AWG. For a continuous AWG, there are no well-defined `channels', however, we can use the 3-dB width of the AWG transmission response at a given location on the FPR as an imaginary limit to define channels. Our goal is to examine whether such definition of channels is invertible to reliably extract the spectrum and what spectral resolution can be achieved using this approach.

\begin{figure}[h!]
\centering
        \includegraphics[width=0.95\textwidth]{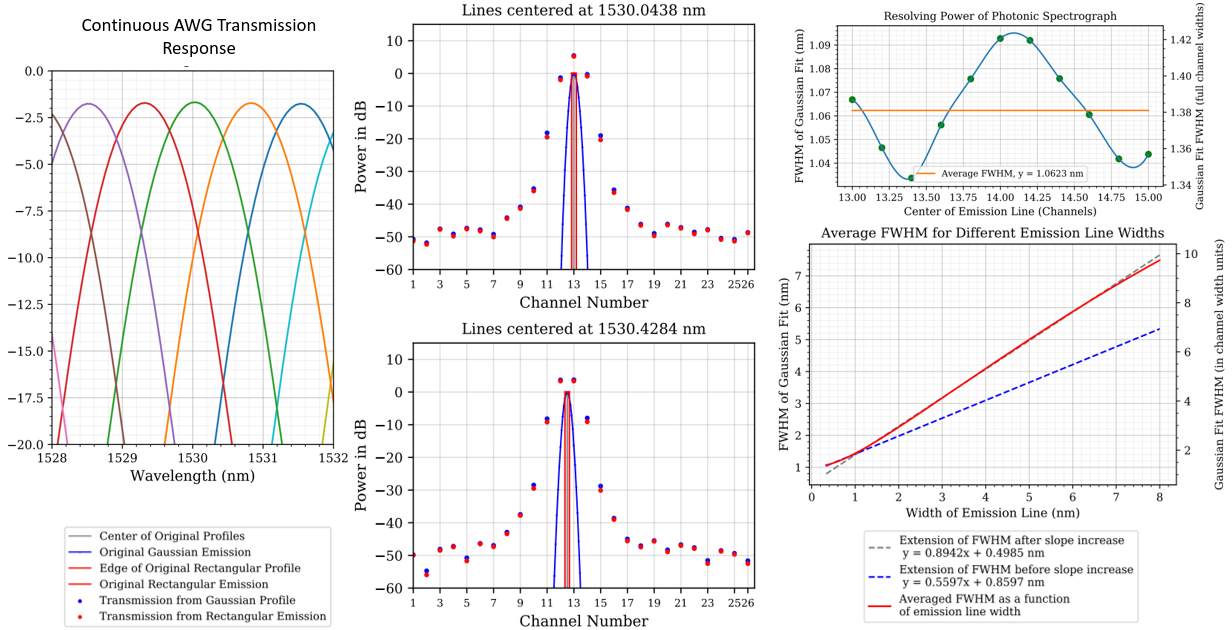}
        \caption{
        \textbf{Left panel:} The transmission response of the continuous AWG by using the 3-dB width of the profiles for defining a spectral `channel'. 
        \textbf{Center panel:} The LSF derived using rectangular (red points) and Gaussian (blue points) emission lines of FWHM = 0.31 nm (i.e. unresolved lines). The LSF is derived for channel 13 (the central channel) and at 50\% offset (halfway between channels 12 and 13). The response is similar for rectangular and Gaussian lines, showing that the LSF is independent of the shape of the input emission line.  
        \textbf{Upper-right panel:} Similar to Fig. \ref{fig:LSF_offset_discrete}, the FWHM of the LSF was obtained at different channel offsets. The FWHM only varies within 5\% as a function of the wavelength offset of the input emission line.    
        \textbf{Lower-right panel:} Similar to Fig. \ref{fig:FWHM_vs_em_width_discrete}, this panel shows the FWHM of the observed spectral profile as a function of the FWHM of the input emission line for 2200 different emission line widths. It is clear that the scaling is linear when the line is resolved (FWHM $>$ 0.8 nm).   
        } 
        \label{fig:FWHM_vs_em_width_continuous}
\end{figure}

Similar to the discrete AWG (Sec \ref{subsec:discrete_AWG}), the transmission response of the AWG, as shown in the leftmost panel of Fig. \ref{fig:FWHM_vs_em_width_continuous}, was obtained by simulating the AWG with beam propagation method using the Rsoft software \cite{BeamPROP}. We then used the 0.31 nm 
as the  width of rectangular emission line and FWHM of Gaussian line (same as Sec \ref{subsec:discrete_AWG}) to ensure that the line remains unresolved. In the central panel Fig. \ref{fig:FWHM_vs_em_width_continuous}, we show the AWG power distribution (and hence, the LSF) corresponding to emission line at the peak wavelength of the central channel (channel 13 at 1530 nm) as well as at an offset of  50\% (1530.43 nm) between channels 13 and 12.  By fitting a Gaussian to the resulting AWG power distribution, we obtained the FWHM of the LSFs. The FWHM for various offset values are shown in the upper-right panel of Fig. \ref{fig:FWHM_vs_em_width_continuous}. It can be clearly seen that while the average FWHM is around 1.06 nm, the FWHM varies  by only $\sim$5\% within the range of 2 channels, depending on the offset from the channel peak wavelength. Thus, we conclude that the AWG LSF is only mildly ($\sim 5\%$) wavelength-dependent.  Note that the average FWHM here is slightly smaller than that for the discrete AWG simply due to the higher number of `effective channels' available for fitting the Gaussian.

Further, in the lower-left panel of Fig. \ref{fig:FWHM_vs_em_width_continuous}, we plot the observed line profile FWHM against the input emission line FWHM for 2200 distinct emission lines (similar to Fig. \ref{fig:FWHM_vs_em_width_discrete}). Again, it is clear from the figure that the observed width scales linearly with the width of the source emission line, which confirms that there are no non-linear artifacts when we use the  AWG as a spectrograph, as expected. The small non-linearity is only seen when the emission line is unresolved (FWHM $<$ 0.8 nm), which is expected in a spectrograph. 

\vspace{-1ex}

\subsection{Full Width at Half Maximum (FWHM) of LSF and Resolving Power}\label{subsec:FWHM}
\vspace{1ex}

\paragraph{Ideal resolution of a discrete AWG}
The FWHM of the LSF as a function of emission line width was derived by fitting a Gaussian to the LSF of the transmission through the photonic spectrograph from different simulated emission lines, examples of which are shown in Figs. \ref{fig:LSF_peaks_discrete}-\ref{fig:FWHM_vs_em_width_continuous}. Note that the Gaussian fits in Figs. \ref{fig:LSF_offset_discrete} and \ref{fig:FWHM_vs_em_width_continuous}  closely approximate the core of the LSF but become less accurate towards the wings of the LSF. This is because the LSF is not exactly Gaussian and its shape changes (albeit slightly) with wavelength. 

\par
The FWHM of the line spread function is not necessarily the best way to define the spectral resolution of the spectrograph, since the FWHM of the LSF can be difficult to compare for spectrographs with different shapes of the LSF \cite{robertson2013quantifying}. The spectral resolution is the minimum separation between two spectral features for them to both be individually resolved by the spectrograph. The FWHM of the LSF is only one part of a calculation for a more consistent calculation of the resolving power that can then be used to compare two spectrographs irrespective of the shape of their LSFs \cite{robertson2013quantifying}. This is non-trivial to calculate for the photonic spectrograph since the shape of the LSF varies with wavelength (as seen in Fig. \ref{fig:LSF_offset_discrete}). To examine the dependence on the shape of the LSF, we have calculated below the spectral resolving power assuming a sinc$^2$ profile and then again assuming a Gaussian profile for the LSF of both the discrete and continuous AWGs.

The generalized equation for converting the FWHM of the LSF to the resolving power the spectrograph is described in Robertson et al. 2013 \cite{robertson2013quantifying} as follows:

\begin{equation}\label{Eqn:Resolution}
R = \frac{\lambda}{\mathrm{\beta ~ FWHM_{LSF}}}
\end{equation}

where $\beta$ is the parameter dependent on the shape of the LSF as follows (see Robertson etal. 2013 \cite{robertson2013quantifying} for detailed derivation):

\begin{equation}\label{Eqn:beta}
    \mathrm{\beta = \frac{1.3809 ~ Z^{1 /3}  EW^{2/3}}{FWHM_{LSF}}}
\end{equation}

where, Z and equivalent-width (EW) are the parameters dependent on the LSF shape. For sinc$^2$ LSF, they are given by\cite{robertson2013quantifying}:

\begin{equation*}
    \mathrm{Z_{sinc^2} = 0.4289~FWHM_{LSF}}
\end{equation*}
\begin{equation*}
    \mathrm{EW_{sinc^2} = FWHM_{LSF}/0.8859}
\end{equation*}
\begin{equation}\label{eqn:beta_sinc2}
    \mathrm{thus~using~Eqn~ \ref{Eqn:beta}:~\beta_{sinc^2} = 1.1289}
\end{equation}

The same parameters for Gaussian LSF are given by \cite{robertson2013quantifying}: 
\begin{equation*}
    \mathrm{Z_{Gaussian} = \frac{FWHM_{LSF}}{\sqrt{2\pi log_{e}(2)}}}
\end{equation*}
\begin{equation*}
    \mathrm{EW_{Gaussian} = FWHM_{LSF}\sqrt{\frac{\pi}{4log_{e}(2)}}}
\end{equation*}
\begin{equation}\label{eqn:beta_gaussian}
    \mathrm{thus~using~Eqn~ \ref{Eqn:beta}:~\beta_{Gaussian} = 1.1265}
\end{equation}

Assuming  a sinc$^2$ profile of the LSF of the discrete AWG and using the average FWHM of the LSF from Fig. \ref{fig:LSF_offset_discrete} (= 1.189 nm), and equations  \ref{Eqn:Resolution} and \ref{eqn:beta_sinc2}, we can deduce the spectral resolving power (Rayleigh criterion) of the photonic spectrograph as:

\begin{equation*}
    R\mathrm{_{Discrete, sinc^{2}}} = \frac{1530.0 nm}{1.1289 \times 1.1899 nm} = 1138.9
\end{equation*}

Similarly, assuming a Gaussian profile for the LSF of the discrete AWG, and using  using equations  \ref{Eqn:Resolution} and \ref{eqn:beta_gaussian}, we get 

\begin{equation*}
    R\mathrm{_{Discrete, Gaussian}} = \frac{1530.0 nm}{1.1265 \times 1.1899 nm} = 1141.4
\end{equation*}

Thus, we should expect a resolving power $R~\sim$ 1140 for the discrete AWG considered here. Similarly, for the continuous AWG, we can use the average FWHM of the LSF from Fig. \ref{fig:FWHM_vs_em_width_continuous} (= 1.0623 nm), and equations  \ref{Eqn:Resolution}, \ref{eqn:beta_sinc2}, and \ref{eqn:beta_gaussian}, we can deduce the spectral resolving power of the continuous AWG as: 

\begin{equation*}
    R\mathrm{_{Discrete, sinc^{2}}} = \frac{1530.0 nm}{1.1289 \times 1.0623 nm} = 1275.7
\end{equation*}

\begin{equation*}
    R\mathrm{_{Continuous, Gaussian}} = \frac{1530.0 nm}{1.1265 \times 1.0623 nm} = 1278.5
\end{equation*}

From this analysis, we conclude that we should expect a resolving power $R~\sim$ 1277 for the continuous AWG considered here, which is larger than the discrete AWG by around 10\%. In addition, we confirm from this analysis that both Gaussian or sinc$^2$ profiles can be used to as the LSF profile and they both give a consistent resolving power.

The next step is to validate this
resolving power by illuminating the discrete and continuous AWGs with a reference spectrum in simulation and then extract an inferred spectrum using the observed power at the end of the AWGs (and a calibration matrix). The goal would be to check if the spectrum can indeed be extracted at the resolution anticipated from the analysis above.


\section{SPECTRAL EXTRACTION}
The AWG spectral response at any output channel (or output waveguide) has a non-zero value at all wavelengths and potentially low-level side lobes at wavelengths other than the central wavelength of that particular channel. Therefore, the AWG spectral extraction cannot simply be done by one-to-one correspondence between the peak wavelength of the channel and the power measured in that particular channel. A reliable and more accurate AWG spectral extraction requires the knowledge of the transmission response of the AWG output channels. The relationship between the AWG response, the source spectrum, and the measured power in AWG output channels can be summarized in a matrix form as follows.

\begin{equation}\label{eqn:AWG_Spec_Construction}
\textbf{A}_{\mathrm{N_{chan} \times N_{resp}}}\cdot\textbf{S}_{\mathrm{N_{resp} \times 1}} = \textbf{P}_{\mathrm{N_{chan} \times 1}}
\end{equation}

The AWG response matrix (\textbf{A}, size: $\mathrm{N_{chan} \times N_{resp}}$) is a direct representation of the transmission response shown in Fig. \ref{fig:figure1} and can be obtained at high-resolution (say, \textbf{$\mathrm{N_{resp}}$} resolution elements, irrespective of the AWG resolution) using a tunable laser source or by using a combined setup of a broadband source and an optical spectrum analyzer \cite{gatkine2021chip}. The observed AWG power in each channel (\textbf{P}) can therefore be expressed by multiplying the AWG transmission response for that channel (at highest resolution) with the source spectrum (at the same resolution, represented by \textbf{S}). Thus, the number of rows in the AWG response matrix is equal to the number of AWG channels (\textbf{$\mathrm{N_{chan}}$}), and the number of columns is equal to the number of resolution elements (\textbf{$\mathrm{N_{resp}}$}) for which the AWG transmission response of the channel was measured.  \vspace{-2ex}

\paragraph{Reference spectrum:} We obtained the near-IR absorption spectrum of CO molecular species at a temperature of 296 K and pressure of 1 atm, over a column length of 1 km using HITRAN \cite{kochanov2016hitran}. We use this spectrum as a reference for testing the performance of the AWG as a spectrograph. For consistency, the spectrum was shifted in the wavelength domain to fall within the 1520$-$1540 nm band, which is used throughout this paper to examine the AWG characteristics. Given the known AWG response for the discrete and continuous AWG, we use Eqn. \ref{eqn:AWG_Spec_Construction} to derive the power measured in each spectral channel of the AWG. The calibration matrix derived in the next subsection is then used to recover the original spectrum (at a low resolution) given the knowledge of the AWG transmission and the power measured in the AWG spectral channels. 

\subsection{Calibration Matrix}
The goal here is to compare the extracted spectrum from the discrete and continuous AWGs of R = 1000 with the original  source/reference spectrum binned at an R = 1000. In order to perform the spectral extraction, the matrix \textbf{A} (size: $\mathrm{N_{chan} \times N_{resp}}$) needs to be inverted. Given that it is a rectangular matrix ($\mathrm{N_{resp} \gg N_{chan}}$), the simplest approach is to bin the AWG transmission response to reduce the number of columns of \textbf{A} from $\mathrm{N_{resp}}$ to $\mathrm{N_{chan}}$ and thus, approximate matrix  \textbf{A} ($\mathrm{N_{chan} \times N_{resp}}$) to $\bm{\mathrm{\tilde{A}}}$ ($\mathrm{N_{chan} \times N_{chan}}$). 
Alternatively, a pseudo-inverse approach \cite{strang1988linear} could be used to invert the rectangular matrix \textbf{A} using its singular value decomposition, but the solution  is unstable and degenerate, which makes it unsuitable for spectral extraction. Hence, we use the binned square matrix $\bm{\mathrm{\tilde{A}}}$ for spectral extraction. The approximate source spectrum as extracted from the AWG is given by pre-multiplying the AWG power (\textbf{P}) with  $\bm{\mathrm{\tilde{A}^{-1}}}$ as follows:

\begin{equation}\label{eqn:AWG_Spec_Reconstruction}
\bm{\mathrm{\tilde{S}}}_{\mathrm{N_{chan} \times 1}} = 
(\bm{\mathrm{\tilde{A}}}_{\mathrm{N_{chan}\times N_{chan}}})^{-1} \cdot \textbf{P}_{\mathrm{N_{chan} \times 1}} 
\end{equation}

\subsection{Results}
\begin{figure}[h!]
\centering
\includegraphics[width=1.0\textwidth]{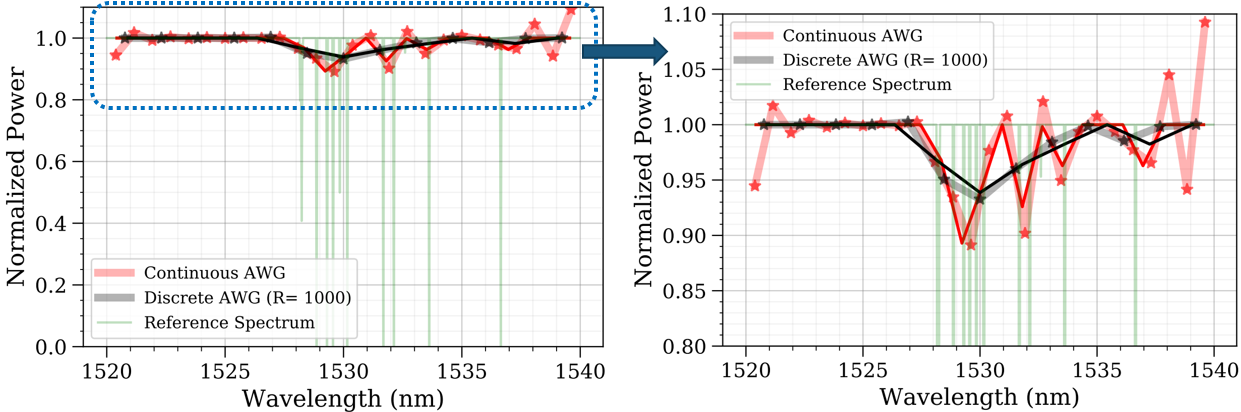}
\caption{\label{Fig:AWG_spec_extraction}AWG spectrum extraction \textbf{Left:} The original high-resolution reference spectrum of the molecular species is shown in green. The solid black line indicates the ideal binned spectrum of the reference spectrum to an R = 1000. The faint black trace and black stars show the extracted reference spectrum if observed with a \textbf{discrete AWG} of R = 1000. The AWG extracted spectrum and the binned reference spectrum are consistent with each other, suggesting a faithful reconstruction. Similarly, the solid red trace shows the reference spectrum binned to an effective spectral resolution of R = 2000. The faint red trace shows the extracted reference spectrum if observed with a \textbf{continuous AWG} of R = 1000. Given the agreement between the continuous AWG and the binned spectrum of R = 2000, is clear that the continuous AWG of R = 1000 has an effective spectral resolution of R $\sim$ 2000. \textbf{Right:} The zoomed-in version of the AWG spectral extraction on the left. }\hfill
\label{fig:AWG_Spec_Extraction}
\end{figure}

Using the spectral extraction method, we extracted the approximate source spectrum for a discrete and continuous AWG of R = 1000 (note that R = $\lambda/\delta\lambda$ with $\delta\lambda$ = channel spacing of the discrete AWG). These results are summarized in Fig. \ref{fig:AWG_Spec_Extraction}. The solid black line indicates the ideal binned spectrum of the reference spectrum to an R = 1000. The faint black trace and black stars show the extracted reference spectrum if observed with a \textbf{discrete AWG} of R = 1000. The AWG extracted spectrum and the binned reference spectrum are consistent with each other, suggesting a faithful reconstruction. Similarly, the solid red trace shows the reference spectrum binned to an effective spectral resolution of R = 2000. The faint red trace and red stars show the extracted reference spectrum if observed with the \textbf{continuous AWG}. Given the agreement between the continuous AWG and the binned spectrum of R = 2000, is clear that the continuous AWG (i.e. the AWG obtained after removing the output waveguides of a discrete AWG of resolution 1000) has an effective spectral resolution of R $\sim$ 2000.

With this simple spectral extraction method, we are able to achieve a spectral resolution R =  $\frac{\lambda}{\delta\lambda} \approx 1,000$ for the discrete AWG and an R = 2000 for the continuous AWG. This is clearly higher than the expected resolution for continuous AWG calculated from its FWHM ($\sim 1275$). This is possible because the AWG transmission response (Fig. \ref{fig:FWHM_vs_em_width_continuous} left panel) can be measured at very high resolution and the higher number of `effective channels' in the continuous AWG (as opposed to the discrete AWG)  allow the inversion of the calibration matrix at a finer resolution. We do see additional outlier structures at the edges of the spectral order for the continuous AWG, which is primarily due to the high overlap between the `effective channels' of a continuous AWG, which can potentially be mitigated using more sophisticated algorithms.   

\subsection{Future Work}

In the future, we plan to work on a more sophisticated algorithm to minimize the stray features in the continuous AWG. Further, we plan to extend this analysis to high-resolution AWGs (R $>$ 10,000) in the future, since  the high-resolution spectroscopy is key for exoplanet atmosphere characterization using instruments such as KPIC. This will allow us to evaluate the performance of compact,  high-resolution AWG spectrographs on a KPIC-like setup for investigating exoplanet atmospheres.

\acknowledgments 
 
M. Perez acknowledges the support from Caltech Summer Undergraduate Research Fellowship (SURF) program and the funding provided by the Flintridge Foundation for his work in the summer of 2021. P. Gatkine was supported by NASA Hubble Fellowship program as well as David \& Ellen Lee Fellowship at Caltech. This work was supported by the Wilf Family Discovery Fund in Space and Planetary Science, funded by the Wilf Family Foundation. This research was carried out at  the California Institute of Technology and the Jet Propulsion Laboratory under a contract with the National Aeronautics and Space Administration (NASA) and funded through the President's and Director's  Research $\&$ Development Fund Program.  

\bibliographystyle{spiebib} 
\bibliography{report} 

\end{document}